# Numeric modeling of synchronous laser pulsing and voltage pulsing field evaporation

L. Zhao[1], A. Normand, J. Houard, I. Blum, F. Delaroche, F. Vurpillot

Normandie Univ, UNIROUEN, INSA Rouen, CNRS, GPM, 76000 Rouen,

**Abstract:** We have recently proposed an atom probe design based on a femtosecond time-resolved pump-probe setup. This setup unlocks the limitation of voltage pulsed mode atom probe thanks to the occurrence of local photoconductive switching effect[2]. In this paper, we have used a numerical model to simulate the field evaporation process triggered by the synchronous two pulses. The model takes into account the local photoconductive effect and the temperature rise caused by the laser application and the voltage pulse distortion due to the RC effect.

## 1. Modeling

Matlab program is used to simulate the pump-probe field evaporation which mainly consists of two basic models: the voltage pulse model and laser pulse model (tip temperature evaluation given in manuscript). Considering the delay ($\delta t$) is zero when the superimposition of the laser pulse with the beginning of the HV pulse plateau, which approximately corresponds to a half path-length of optical delay line. According to an Arrhenius law (given in the manuscript), the ion flux can be simply computed as a function of the delay ($\delta t$) between two models.

A mathematical approximation of the measured pulse shape[1] was then used in the simulation, expressed as:

$$V_p(t) = -V_0 \exp(-\left(\frac{t}{2\sigma}\right)^{2n}) \qquad \text{Eq. 1}$$

---

[1] Corresponding author: lu.zhao@univ-rouen.fr
[2] This paper provides the supplemental modeling information for the paper "Nanoscale photoconductive switching effect applied to atom probe tomography".

where $V_0$ is the maximum amplitude of the pulse. With the pulse shape indicator $n = 9$ and $\sigma = 175$, a square pulse model with duration (FWHM) about 350ps was produced.

### 1.1 For metal case

Although the double pulses apply to the sample, there are not free carriers generated after laser pulses impact on the tungsten. A sharp temperature rise on the tip caused by the absorption of a part of laser energy then it cools down to the base temperature. This temperature evaluation is given in manuscript. We only model the temporal delay ($\delta t$) as mathematical function translation, written as:

$$T(t) = T_0 + \frac{T_{rise}}{\sqrt{1 + 2\frac{t - \delta t}{\tau_{cooling}}}} \qquad \text{Eq. 2}$$

with $t_{cooling} = \omega^2/\alpha$.

where $T_0$ the base temperature, $T_{rise}$ the temperature rise caused by the absorption of laser energy, $\omega$ the size of the heated zone at the apex of the sample and $\alpha$ the thermal diffusivity of the sample. It is evident that the size of the heated zone is greater, the cooling time is longer.

In order to simulate the ion flux, the DC voltage ($V_{DC}$) was adjusted at 85% of the voltage required to DC field evaporate the sample ($V_e$), and the pulse ratio ($V_p/V_0$) equals to 5% to match the experimental conditions. Here we have taken Eq.4 from the manuscript to compute ion flux. Without laser pulse excitation, the calculated ion flux is negligible ($\sim 10^{-11}$ atom/pulse) with $\nu = K_b T/h \sim 2 * 10^{12}$ Hz and the activation energy ($Q_0$) of 2eV. Taking the laser pulse model described by Eq. 2 to describe the temperature of the tip, which had a sharp rise from 90K to 200K. According to the previous study[2], the cooling time of tungsten in the simulation was configured of 650ps. Supplementary Figure 1 gives the temperature evaluation of sample when the laser pulse impacts on the tip.

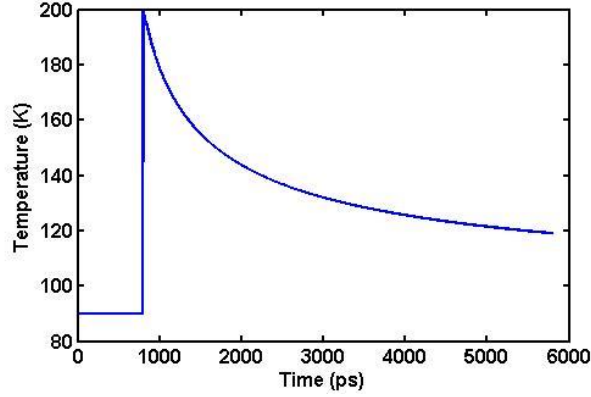

**Supplementary Figure 1: Temperature evaluation after the laser pulse (occurred at 800ps) applied to the tip of the sample.**

For each step of optical delay, it is possible to calculate the ion flux which is illustrated in manuscript (Figure 2(a)).

### 1.2 For semiconductor case

#### 1.2.i Photoelectric effect

The key physical mechanism of photoelectric effect is the energy exchange between electron energy and photon energy ($E_{photon}$) which is defined as:

$$E_{photon} = hv = \frac{hc}{\lambda} \qquad \text{Eq. 3}$$

Where $h$ is the Planck constant, $v$ stands for the frequency, $\lambda$ the wavelength of the light and c is the speed of light. The photoexcitation is only if the photon energy at least as large as the bandgap energy of semiconductor. Supplementary Figure 2 shows the intrinsic/extrinsic photoexcitation effect inside a semiconductor. While the light is taken off, the conduction electrons will recombine with holes to release energy in form of light or thermal energy and the concentration of the carriers could be expressed by[3]

$$\Delta n(t) = n_0 \exp(-t/\tau) \qquad \text{Eq. 4}$$

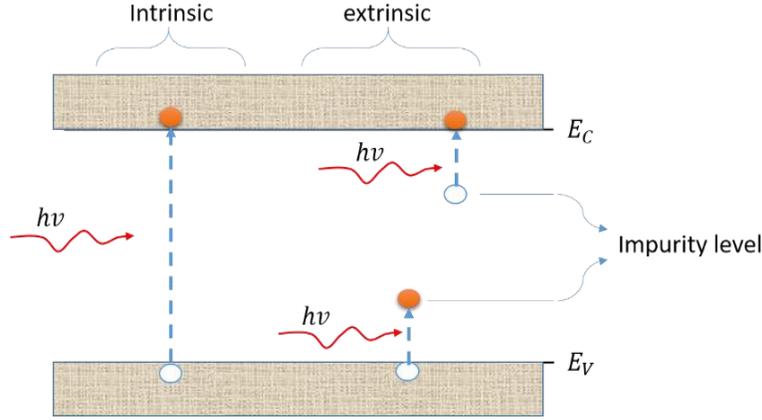

Supplementary Figure 2: Photoexcitation from band-to-band for intrinsic semiconductor and photoexcitation between impurity level and band for extrinsic semiconductor[3].

where $n_0$ is the carrier concentration when the light is taken off and $\tau$ is the carrier lifetime.

When the device is under illumination, the conductivity of the device is given by

$$\sigma = q(n_i + \Delta n)(u_n + u_p) = qn_i(u_n + u_p) + q\Delta n(u_n + u_p) \qquad \text{Eq. 5}$$

We notice that the conductivity thus can be written as $\sigma = \sigma_i + \Delta\sigma$, where $\Delta\sigma = q\Delta n(u_n + u_p)$ is the conductivity induced by the generation of the free carries ($\Delta n$).

For a device with a given length $l$ and a cross-sectional area $A$, the resistance $R_s$ can be computed as

$$R_s = \frac{\rho l}{A} = \frac{l}{\sigma A} \qquad \text{Eq. 6}$$

Substituting for $\sigma$ gives

$$R_s = R_i \left(\frac{1}{1 + \frac{\Delta\sigma}{\sigma_i}}\right) \qquad \text{Eq. 7}$$

where $R_i = \frac{l}{\sigma_i A}$ is the resistance of device under steady-state condition.

With $\frac{\Delta\sigma}{\sigma_i} \gg 1$, we can simplify Eq. 7 and the resistance of device could be rewritten as

$$R_s = R_i \frac{\sigma_i}{\Delta\sigma} \qquad \text{Eq. 8}$$

Note that the resistor of $R_s$ is generally at least $10^4$ times less than dark resistor $R_i$ in photoconductive switch application.

The resistance drop enables a semiconductor device to act as closing switch[4], [5], indeed, the response of photoexcitation occurs in 1ps order, i.e. the resistance drop of the device could follow the laser pulse variation.

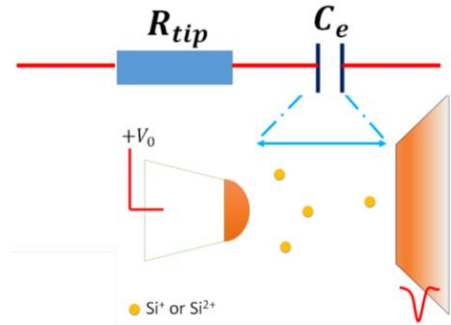

**Supplementary Figure 3: Schematic diagram of tip-microelectrode setup in atom probe, for instance, intrinsic silicon sample. Considering the whole environment surrounding the tip (chamber, local electrode...), this setup can be simply represented by RC low-pass filter.**

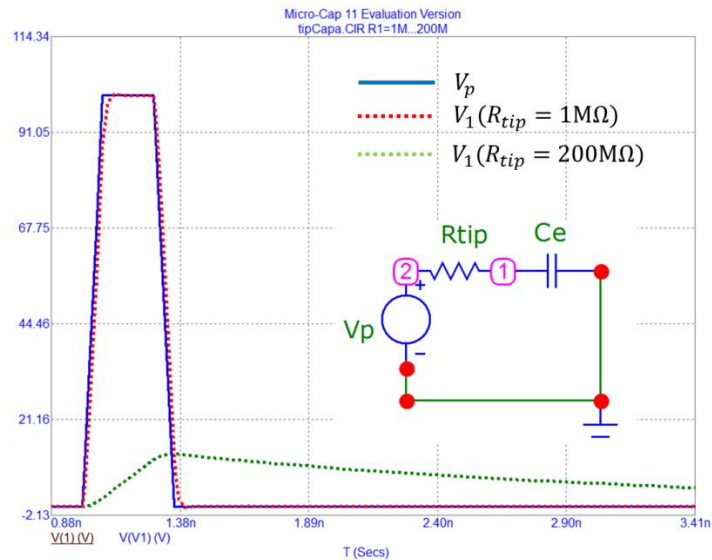

**Supplementary Figure 4: Lumped simulation model for RC effect: the voltage pulse profile between the $R_{tip}$ and $C_e$ depends on the cut-off frequency of the circuit. This effect may enlarge the pulse width and significantly attenuate the pulse amplitude (green dot curve). Decreasing the resistance of $R_{tip}$ allows transmitting short voltage pulse with slight degradation (red dot curve). The blue curve shows the input signal $V_p$.**

The sample in atom probe experiment is typically prepared in the form of a needle-shaped tip. To be simplified, we consider the tip is hemisphere on a cylinder with a radius around R=50nm. Even the laser beam from one side of the tip, perpendicular to the sample axis, the generation of free carriers could be homogeny inside the tip, since the absorption depth d is about 100nm at a wavelength of 400nm for silicon[6]. This variation of resistance thus can be described by a simple 1D model as:

$$R(t) = R_{min} + R_{max}\left(1 - \exp(-(t - \delta t)/t_g)\right) \qquad \text{Eq. 9}$$

where $t_g$ being the life-time of free carriers, $R_{max}$ the resistance of tip without illumination and $R_{min}$ the resistance when the laser pulse applied to the tip. This model was taken into account being independent to the temperature variation induced by laser pulse. Using a femtosecond laser pulse with relatively low energy to create the free carries, this lifetime should to be much shorter than that in photoconductive switch application[7].

And moreover, the tip-electrode geometry in HV pulsed atom probe is typically studied as RC series circuit[8] dedicated in Supplementary Figure 3. The RC effect can enlarge the voltage pulse width and significantly decreases the amplitude, an example shown in Supplementary Figure 4.

The resistance of the intrinsic silicon sample at low temperature (few K) might be greater than $1G\Omega$, so it can be considered as a quasi-insulator. With an additional effect: RC effect with the tip capacitance is extremely small (in the range of sub-fF domain), no effective signal can be transmitted, *i.e. the voltage pulse is unable to produce any field evaporation whatever the amplitude of the voltage pulses*. That is why the voltage pulse atom probe (Ex. LEAP) only used to analyze the material with low resistivity (much less than $10\Omega.cm$).

In the numerical simulation of semiconductor case, the DC voltage was set about 90% of the required field evaporation voltage and the laser energy was adjusted half less than traditional laser assisted atom probe, then the triggering evaporation flux is in charge of the voltage pulse. The base temperature of the sample was also set at 90K to replicate the experimental condition, and the sharp rising temperature is about 40K (because less light energy applied to the tip) which is much less than that of tungsten, the sample temperature was thus cooled down from the maximum temperature(130K) with $t_{cooling} = $ 2ps[9]. Note that there was an ion flux close to zero when the pulses were not synchronized. As simulated in the metal case, the voltage pulse will be scanned by the temperature model of laser, the effective voltage pulse ($V_c$) experienced by surface atoms will be numerically recalculated by solving the differential equations describing the RC circuit:

$$\frac{V_p(t) - V_c(t)}{R(t)} = C\frac{dV_c(t)}{dt} \qquad \text{Eq. 10}$$

where $C$ is the equivalent capacitance between the tip and the electrode. The effective voltage can be computed by substituting $V_p(t)$. Supplementary Figure 5 gives the voltage pulse profiles when the laser pulse applied to the sample with different delays, i.e. the laser pulse applied to the sample from before to after the voltage pulse application. To simulate the ion flux, replacing $V_p(t)$ by $V_c(t)$ in the Eq. 4 given in the manuscript, the ion flux as a function of delay was obtained. The calculated results give a good agreement to experimentally observed ion flux when we fix the fast cooling time $t_{cooling} = 2ps$ and adjust the short recombination time $t_g$ to 100ps and the tip resistance to reducing 200 times under the light illumination.

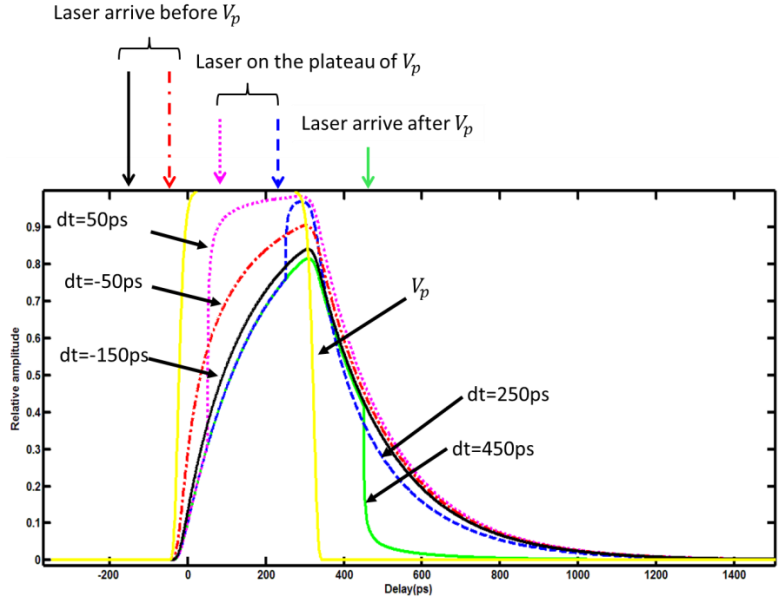

**Supplementary Figure 5:** Actual voltage profile experienced by tip surface atoms with different delay ($\delta t$): Laser arriving before voltage pulse to the sample ($\delta t < 0$), the pulse represents a typically charging and discharging process with strongly reduced amplitude due to RC effect. When the laser pulse superimpose with the plateau of $V_p$ ($\delta t = 0 - 350ps$), voltage pulse is divided into two: before application of the laser pulse, the voltage pulse follows the charging edge; after application of the laser pulse, the voltage pulse approach the original signal ($V_p$) until the end of voltage pulse plateau, and then decay decreased as the discharging process. Laser pulse arriving after the voltage pulse finished ($\delta t > 350ps$), the output signal mainly follows the charging process, and however, the discharging decay will be cut at the moment of the application of laser pulse (at $\delta t = 450ps$).

# Reference

[1]  L. Zhao, a. Normand, F. Delaroche, B. Ravelo, and F. Vurpillot, "Pulse shaping optimization for improving atom probe tomography," *Int. J. Mass Spectrom.*, vol. 386, pp. 47–53, Jul. 2015.

[2]  F. Vurpillot, J. Houard, a Vella, and B. Deconihout, "Thermal response of a field emitter subjected to ultra-fast laser illumination," *J. Phys. D. Appl. Phys.*, vol. 42, no. 12, p. 125502, 2009.

[3]  K. K. N. Simon M. Sze, *Physics of Semiconductor Devices*, 3rd ed. Wiley.

[4]  D. H. Auston, "Picosecond optoelectronic switching and gating in silicon," *Appl. Phys. Lett.*, vol. 101, pp. 24–27, 1975.

[5]  C. H. Lee, "Picosecond optoelectronic switching in GaAs ," *Appl. Phys. Lett.*, vol. 30, no. 2, p. 84, 1977.

[6]  K. S.KELKAR, "Silicon Carbide As a Photo Conductive Swtich Material for High," 2006.

[7]  C. H. Lee, "Optical control of semiconductor closing and opening switches," *IEEE Trans. Electron Devices*, vol. 37, no. 12 !I, pp. 2426–2438, 1990.

[8]  M. Gilbert, F. Vurpillot, A. Vella, H. Bernas, and B. Deconihout, "Some aspects of the silicon behaviour under femtosecond pulsed laser field evaporation.," *Ultramicroscopy*, vol. 107, no. 9, pp. 767–72, 2007.

[9]  A. Vella, E. P. Silaeva, J. Houard, T. E. Itina, and B. Deconihout, "Probing the thermal response of a silicon field emitter by ultra-fast Laser Assisted Atom Probe Tomography," *Ann. Phys.*, vol. 525, no. 1–2, pp. L1–L5, 2013.